\documentclass[aps,prd,reprint,floatfix,nofootinbib,superscriptaddress]{revtex4-2}
\usepackage{amsmath,amssymb,mathtools,bm,graphicx,booktabs,xcolor,microtype,hyperref}
\hypersetup{colorlinks=true,linkcolor=blue,citecolor=blue,urlcolor=blue}
\newcommand{\dd}{\mathrm d}\newcommand{\ee}{\mathrm e}
\newcommand{\eps}{\varepsilon}
\allowdisplaybreaks[2]
\graphicspath{{figures/}}
\begin{document}
\title{Distinguishing Spacetime Geometries beyond the Leading Power Law in Strong-Deflection Lensing near Degenerate Photon Spheres}

\author{Yang Liu}
\email{yliu1024@nuaa.edu.cn}
\affiliation{Center for the Cross-disciplinary Research of Space Science and Quantum-technologies (CROSS-Q), College of Physics, Nanjing University of Aeronautics and Astronautics, \\29 Jiangjun Road, Nanjing City, Jiangsu Province 211106, China}

\author{Xin Zhao}
\email{zhaox923@nuaa.edu.cn}
\affiliation{Center for the Cross-disciplinary Research of Space Science and Quantum-technologies (CROSS-Q), College of Physics, Nanjing University of Aeronautics and Astronautics, \\29 Jiangjun Road, Nanjing City, Jiangsu Province 211106, China}
 
\author{Chen-Hao Wu}
\email{chenhao\_wu@nuaa.edu.cn}
\affiliation{Center for the Cross-disciplinary Research of Space Science and Quantum-technologies (CROSS-Q), College of Physics, Nanjing University of Aeronautics and Astronautics, \\29 Jiangjun Road, Nanjing City, Jiangsu Province 211106, China}

\begin{abstract}
Black hole imaging by the Event Horizon Telescope and future high-resolution observations of photon rings offer new opportunities to test black hole geometries and distinguish black holes from their mimickers.
Light propagation near circular photon orbits is central to these investigations.
The merger of two unstable circular photon orbits and an intermediate stable orbit can produce a quartic maximum of the null effective potential.
Near such an orbit, the deflection angle exhibits a leading power-law divergence.
Exploring divergences beyond the leading power law can reveal additional geometric information for distinguishing spacetimes.
For rays that turn outside the quartic orbit, we derive a strong-deflection expansion in static, spherically symmetric, and asymptotically flat spacetimes.
The deflection angle can exhibit both power-law and logarithmic divergences.
We apply the expansion to the Simpson--Visser wormhole, a hairy Schwarzschild black hole, and a black hole with a compact Hernquist-type matter halo.
The logarithmic term produces a subleading correction to high-order image positions.
The ratio of the logarithmic and leading image-position coefficients determines the logarithmic deflection coefficient independently of the source and observer locations.
We demonstrate that the logarithmic correction can distinguish geometries with the same critical impact parameter and leading deflection coefficient.
These results provide an additional geometric constraint for testing candidate spacetime metrics.
\end{abstract}

\maketitle

\section{Introduction}
\label{sec:introduction}

The images of M87* and Sgr A* obtained by the Event Horizon Telescope have opened a new window for probing spacetime geometry in the strong-gravity regime~\cite{EventHorizonTelescope:2019dse, EventHorizonTelescope:2022wkp}.
These observations have been used to constrain black hole metrics and test alternative compact-object models~\cite{EventHorizonTelescope:2022xqj}.
Future observations with higher angular resolution may probe the structure of photon rings and provide further constraints on spacetime geometry \cite{Johnson:2019ljv,Gralla:2020srx}.
An important step toward these goals is to understand how spacetime geometry affects strong gravitational lensing.

In static, spherically symmetric spacetimes, light rays passing close to an unstable photon sphere can wind around a compact object several times before reaching an observer and forming high-order images~\cite{Claudel2001,Virbhadra2000,Perlick2004Review}.
For returning rays near a nondegenerate unstable photon sphere, the deflection angle diverges logarithmically as the impact parameter $b$ approaches its critical value $b_m$ \cite{Darwin1959,Bozza2002,Eiroa2002RN,Tsukamoto2017SDL}.
The logarithmic coefficient is determined by the geometry near the photon sphere and controls the exponential accumulation of high-order images for a fixed source and observer \cite{Bozza2001,Bozza2002,Bozza2010,Igata2026Local}.
Related work has examined corrections to the logarithmic deflection law in Schwarzschild and Reissner--Nordstr\"om spacetimes \cite{IyerPetters2007,Tsukamoto2022Affine} and strong deflection in spacetimes without event horizons, including those with naked singularities or traversable wormholes~\cite{Shaikh2019Inside, Tsukamoto2021RN,Tsukamoto2016Ellis,Shaikh2019Wormholes, Tsukamoto2021,Tsukamoto2022Retrolensing}.

Some spacetimes contain two unstable circular photon orbits separated by a stable orbit \cite{Guo2023Multiple}.
Such configurations can produce additional sequences of high-order images~\cite{Guo2022Lensing}.
When neighboring circular photon orbits merge, the second radial derivative of the null effective potential vanishes at the merged orbit.
Higher-order terms then govern the leading deflection behavior.
A stable orbit and a neighboring unstable orbit can merge to form a stationary inflection of the potential with a nonzero cubic term.
The corresponding returning rays obey $\Theta\propto\eps^{-1/6}$ at leading order, where $\eps=|b/b_m-1|$ \cite{ChibaKimura2017,Tsukamoto2020}.
The coefficients and finite terms of this law have been studied for rays turning inside or outside the degenerate orbit~\cite{Sasaki2025Marginal, Tsukamoto2026InsideOutside,Igata2026Degenerate}.

When two unstable circular photon orbits merge with the stable orbit between them, the effective potential can have a quartic maximum.
For returning rays approaching this orbit, the leading deflection follows $\Theta\propto\eps^{-1/4}$.
Both cases give a power-law accumulation of high-order images.

Studies of wormhole throats have addressed the leading $\eps^{-1/4}$ divergence, finite contributions to the deflection angle, and lensing observables \cite{Tsukamoto2020DSWormhole,Tsukamoto2021SimpsonVisser, FuZhang2022LQG,ZhangXie2024Wormhole}.
Recent work has revisited the strong-deflection coefficients at reflection-symmetric throats \cite{Tsukamoto2026Z2Throat}.
General formulas for the leading divergence near degenerate photon orbits also include the quartic case~\cite{OvgunPantig2026Degenerate}.

Knowledge of the critical impact parameter and the leading deflection coefficient alone may not suffice to distinguish different spacetime geometries.
We show that the logarithmic contribution distinguishes such geometries by probing local properties not determined by these quantities.
We establish how this additional information can be recovered from high-order image positions, providing a further test of candidate spacetime metrics.

We consider an asymptotically flat, static, and spherically symmetric spacetime.
The rays turn outside a quartic maximum of the null effective potential and return to the same asymptotic region.
For these rays, $b>b_m$, so we use $\eps=b/b_m-1>0$.
We derive the expansion
\[
 \Theta
 =\bar c\eps^{-1/4}+\ell\ln\eps+d+o(1),
 \qquad \eps\to0^+.
\]
Both $\bar c$ and $\ell$ depend on the geometry near the orbit.
The coefficient $\ell$ involves local geometric quantities that do not enter $\bar c$.
The finite term $d$ depends on the full returning path.

We apply the expansion to the Simpson--Visser wormhole~\cite{SimpsonVisser2019}, a hairy Schwarzschild (hSBH) black hole \cite{Ovalle2021}, and a black hole with a compact Hernquist-type matter halo \cite{Cardoso2022Halo,Xavier2023Halo}.
Direct numerical evaluation of the deflection angle confirms the predicted asymptotic behavior in all three examples.

For a fixed source and observer at finite radii, the leading image offsets from the critical circle decrease as $n^{-4}$, where $n$ counts the additional full turns of the ray.
The logarithmic deflection coefficient $\ell$ controls the $n^{-5}\ln n$ correction to these offsets.
Numerical image sequences confirm the predicted asymptotic structure in the three original models.

The ratio of the logarithmic and leading coefficients in the asymptotic image expansion determines $\ell$ independently of the source and observer positions.
A local modification of the halo metric leaves the critical impact parameter and $\bar c$ unchanged while varying $\ell$.
For the modified halo geometries, placing the source and observer at the same fixed positions outside the modified region gives the same critical angle and leading image coefficient, but different logarithmic corrections.
An inferred value of $\ell$ can then be compared with predictions of candidate spacetime metrics, providing an additional constraint on the geometry near the photon orbit.

Section~\ref{sec:deflection} introduces the circular-orbit conditions and the deflection integral and reviews the quadratic and cubic deflection laws.
Section~\ref{sec:critical} derives the general quartic expansion.
Section~\ref{sec:geometry} applies the expansion to three spacetime geometries.
Section~\ref{sec:images} relates high-order image positions to the logarithmic deflection coefficient.
Section~\ref{sec:conclusion} presents our conclusions.
Throughout this paper, we use $G=c=1$.

\section{Photon spheres and light deflection}
\label{sec:deflection}

\subsection{Photon orbits and the deflection angle}
\label{sec:returning_general}
\label{sec:local_orbits}

Consider an asymptotically flat, static, spherically symmetric metric
\begin{equation}
 \dd s^2
 =-A(r)\dd t^2+B(r)\dd r^2
 +C(r)\left(\dd\theta^2+\sin^2\theta\,\dd\phi^2\right).
 \label{eq:general_metric}
\end{equation}
At infinity, $A,B\to1$ and $C/r^2\to1$.

For an equatorial null geodesic, $E=A\dot t$ and $L=C\dot\phi$ are the conserved energy and angular momentum, and $b=L/E>0$ is the impact parameter.
Dots mean derivatives with respect to an affine parameter.
The null effective potential is
\begin{equation}
 V(r)=\frac{A(r)}{C(r)}.
 \label{eq:effective_potential}
\end{equation}

Let $r_0$ be the turning point, where $\dot r=0$, and use a subscript $0$ for evaluation there.
The null condition and this turning-point condition give
\begin{equation}
 \dot r^2
 =\frac{E^2}{AB}\left(1-b^2V(r)\right),
 \qquad
 b^2=\frac{C_0}{A_0}.
 \label{eq:general_path}
\end{equation}

We consider rays that come from infinity, reach a turning point, and return to the same asymptotic end.
We choose $r$ to increase toward that end and take $A,B,C>0$ along these paths.
As the ray moves inward from infinity, it turns at $r_0$ where $V(r_0)=b^{-2}$.
Before the turn, $\dot r^2>0$, so Eq.~\eqref{eq:general_path} requires $V(r)<b^{-2}$ for $r>r_0$ along the path.
When the potential has several barriers, these conditions determine where the incoming ray turns~\cite{Pradhan2026TwoThroat}.

For a photon to remain on a circular orbit, both its radial velocity and acceleration must vanish.
We use the subscript $m$ for quantities evaluated at the circular orbit.
The orbit is at $r=r_m$, with critical impact parameter $b_m$.
The null geodesic equations give
\begin{equation}
 V'_m=0,
 \qquad
 b_m^2=\frac{C_m}{A_m}=\frac{1}{V_m}.
 \label{eq:circular_conditions}
\end{equation}
Primes denote derivatives with respect to $r$.

Circular null orbits are radially unstable at local maxima of $V$ and stable at local minima.
The equation $V'(r)=0$ may have several roots in this region, each corresponding to a circular null orbit.
A stable orbit and a neighboring unstable orbit can merge at a cubic stationary inflection of $V$, where $V'_m=V''_m=0$ and $V^{(3)}_m\ne0$.
We focus on quartic maxima of $V$, which can form when two unstable orbits merge with the stable orbit between them.
At such a maximum, $V'_m=V''_m=V^{(3)}_m=0$ and $V^{(4)}_m<0$.

The azimuthal change $I$ and the deflection angle $\Theta$ are given by~\cite{Bozza2002,Perlick2004}
\begin{equation}
\begin{aligned}
 F(r,r_0)
 &=A_0-C_0\frac{A(r)}{C(r)}
   =C_0[V(r_0)-V(r)],\\
 \bar S(r,r_0)
 &=\frac{2\sqrt{A(r)B(r)}\sqrt{C_0}}{C(r)},\\
 I(r_0)
 &=\int_{r_0}^{\infty}
   \frac{\bar S(r,r_0)}{\sqrt{F(r,r_0)}}\,\dd r,\quad
 \Theta(r_0)
 =I(r_0)-\pi.
\end{aligned}
\label{eq:original}
\end{equation}

\subsection{Logarithmic and power-law deflection}
\label{sec:exact_expansion}

We consider returning rays in the limit $r_0\to r_m^+$, with the metric parameters fixed.
We assume that the contribution from outside a fixed neighborhood of $r_m$ has a finite limit.

We use a radial coordinate regular at $r_m$, with finite positive $A_m,B_m,C_m$.
At fixed $r_0$, we expand $F(r,r_0)$ in powers of $r-r_0$ through order $N$:
\begin{equation}
\begin{aligned}
 F(r,r_0)
 &=\sum_{n=1}^{N}h_{n,0}(r-r_0)^n
   +O(|r-r_0|^{N+1}),\\
 h_{n,0}
 &=-\frac{C_0}{n!}V^{(n)}(r_0),
 \qquad 1\le n\le N.
\end{aligned}
\label{eq:Taylor5}
\end{equation}

At $r_0=r_m$, Eqs.~\eqref{eq:circular_conditions} and~\eqref{eq:Taylor5} give $h_{1,m}=0$.
For the returning rays considered here, the first nonzero term in $F(r,r_m)$ gives the following cases:
\begin{equation}
\begin{array}{@{}c|l@{}}
 \text{quadratic}&h_{1,m}=0,\quad h_{2,m}>0,\\
 \text{cubic}&h_{1,m}=h_{2,m}=0,\quad h_{3,m}>0,\\
 \text{quartic}&h_{1,m}=h_{2,m}=h_{3,m}=0,\quad h_{4,m}>0.
\end{array}
\label{eq:coefficient_orders}
\end{equation}
Since $C_m>0$, each nonzero $h_{n,m}$ has the opposite sign to $V_m^{(n)}$.
The quadratic and quartic cases describe local maxima of $V$, while the cubic case describes a stationary inflection.

For these rays, $b\to b_m^+$.
We define
\begin{equation}
 \eps=\frac{b}{b_m}-1>0.
 \label{eq:impact_offset}
\end{equation}
The subscript $+$ indicates the limit $b\to b_m^+$.

We examine the deflection integral in Eq.~\eqref{eq:original} as $\eps\to0^+$.
The value of $\bar S(r,r_0)$ at the orbit is
\begin{equation}
 S_m=\bar S(r_m,r_m)
     =\frac{2\sqrt{A_mB_m}}{\sqrt{C_m}}.
 \label{eq:prefactor_m}
\end{equation}

For $h_{2,m}>0$, the deflection angle diverges logarithmically~\cite{Darwin1959,Bozza2002, Tsukamoto2017SDL,Igata2026Local,Gao2021}:
\begin{equation}
\begin{aligned}
 \Theta_{2,+}
 &=-\bar a_m\ln\eps+d_++o(1),\\
 \bar a_m
 &=\frac{S_m}{2\sqrt{h_{2,m}}}
  =\sqrt{\frac{2A_mB_m}{A_mC''_m-C_mA''_m}}.
\end{aligned}
\label{eq:ordinary_log}
\end{equation}
The finite term $d_+$ depends on the full returning path.
Appendix~\ref{app:bozza} examines the Schwarzschild residual integral of Ref.~\cite{Bozza2002} as $r_0\to r_m^+$.

When $h_{2,m}=0$ and $h_{3,m}>0$, the deflection follows a power law~\cite{Tsukamoto2020,Tsukamoto2026InsideOutside, Igata2026Degenerate}:
\begin{equation}
\begin{aligned}
 \Theta_{3,+}
 &=\bar c_{3,+}\eps^{-1/6}+d_3+o(1),\\
 \bar c_{3,+}
 &=S_mJ_{3,+}h_{3,m}^{-1/3}(2A_m)^{-1/6}.
\end{aligned}
\label{eq:marginal_coeffs}
\end{equation}
Here $J_{3,+}=B(1/6,1/2)/3$, where $B(x,y)$ is the Euler beta function.
The finite term $d_3$ depends on the full returning path.

Suppose $V$ has two quadratic maxima separated by a minimum.
Both unstable orbits give logarithmic limits under the path conditions above.
The outer maximum can merge with the minimum into a cubic stationary inflection.
The merged orbit then gives the cubic law.
The inner maximum keeps its logarithmic law if it remains quadratic.

If $h_{2,m}=h_{3,m}=0$ and $h_{4,m}>0$, the leading behavior is $\Theta\propto\eps^{-1/4}$.
This law has been obtained at wormhole throats~\cite{Tsukamoto2020DSWormhole, Tsukamoto2021SimpsonVisser} and follows from general results for higher-order degeneracy~\cite{OvgunPantig2026Degenerate}.

For quadratic and cubic photon orbits, only the leading term of the integrand at $r_0=r_m$ has a divergent integral near the orbit.
The quartic case is the first in which a subleading term can also produce a divergence.
The resulting logarithmic contribution carries local geometric information beyond the leading power-law term.

\section{Strong deflection near a quartic photon sphere}
\label{sec:critical}
\label{sec:quartic_cusp}

We derive the strong-deflection expansion near a quartic photon sphere.
We determine how the local geometry enters the logarithmic coefficient, which contains information not fixed by the critical impact parameter and the leading deflection coefficient.
This information can be inferred from high-order image positions and used to constrain candidate spacetime metrics, as shown in Sec.~\ref{sec:images}.

\subsection{Origin of the logarithmic term}
\label{sec:quartic_divergent}

We first set $r_0=r_m$ in the integrand of Eq.~\eqref{eq:original} and examine $\bar S(r,r_m)/\sqrt{F(r,r_m)}$ as $r\to r_m^+$.

In the quadratic and cubic cases, only the leading term of this integrand has a divergent integral.
The remaining terms are $O(1)$ in the quadratic case.
They are $O((r-r_m)^{-1/2})$ in the cubic case.
These terms are integrable near $r_m$.

For a quartic maximum of $V$, the integrand has a leading $(r-r_m)^{-2}$ term.
The next term, proportional to $(r-r_m)^{-1}$, also has a divergent integral if its coefficient is nonzero.
We therefore expand $F(r,r_m)$ through fifth order and $\bar S(r,r_m)$ through first order in $r-r_m$.
With $h_{1,m}=h_{2,m}=h_{3,m}=0$ and $h_{4,m}>0$, Eq.~\eqref{eq:Taylor5} gives
\begin{equation}
 F(r,r_m)
 =h_{4,m}(r-r_m)^4+h_{5,m}(r-r_m)^5
 +O((r-r_m)^6).
\label{eq:quartic_critical_root}
\end{equation}
Expanding the numerator $\bar S(r,r_m)$ about $r=r_m$ gives
\begin{equation}
\begin{aligned}
 \bar S(r,r_m)
 &=S_m+S_{1,m}(r-r_m)+O((r-r_m)^2),\\
 S_{1,m}
 &=\left.\partial_r\bar S(r,r_m)\right|_{r=r_m}
 =S_m\left[
 \frac{A'}{2A}+\frac{B'}{2B}-\frac{C'}{C}
 \right]_m.
\end{aligned}
\label{eq:prefactor_local}
\end{equation}
Combining Eqs.~\eqref{eq:quartic_critical_root} and~\eqref{eq:prefactor_local}, we obtain
\begin{equation}
\begin{aligned}
 \frac{\bar S(r,r_m)}{\sqrt{F(r,r_m)}}
 &=\frac{1}{\sqrt{h_{4,m}}}
 \left[
 \frac{S_m}{(r-r_m)^2}
 +\frac{T_{r,m}}{r-r_m}
 \right]+O(1),\\
 T_{r,m}
 &=S_{1,m}-\frac{S_mh_{5,m}}{2h_{4,m}}.
\end{aligned}
\label{eq:quartic_critical_density}
\end{equation}
The $(r-r_m)^{-2}$ term has an antiderivative proportional to $(r-r_m)^{-1}$, while the $(r-r_m)^{-1}$ term gives a logarithmic divergence when $T_{r,m}\ne0$.
The remaining $O(1)$ terms are integrable near $r_m$.

\subsection{Deflection angle of returning rays}
\label{sec:quartic_finite}

The expansion at $r_0=r_m$ identifies the terms with divergent integrals.
We now consider returning rays with turning points $r_0>r_m$ and determine the divergent and finite terms of $\Theta$ as $r_0\to r_m^+$.

Set $k_0=C_0/C_m$.
Equation~\eqref{eq:original} gives $\bar S(r,r_0)=\sqrt{k_0}\bar S(r,r_m)$.
Using this relation and $A_0/C_0=b^{-2}$, we obtain
\begin{equation}
\begin{gathered}
 \frac{F(r,r_0)}{k_0}
 =F(r,r_m)-F(r_0,r_m),\\
 \frac{\bar S(r,r_0)}{\sqrt{F(r,r_0)}}
 =\frac{\bar S(r,r_m)}
 {\sqrt{F(r,r_m)-F(r_0,r_m)}},\\
 F(r_0,r_m)
 =A_m-C_m\frac{A(r_0)}{C(r_0)}
 =A_m\left[1-\left(\frac{b_m}{b}\right)^2\right].
\end{gathered}
\label{eq:exact_split}
\end{equation}

We next expand $h_{n,0}/k_0$ in powers of $r_0-r_m$.
Differentiating $h_{n,0}/k_0=-C_mV^{(n)}(r_0)/n!$ with respect to $r_0$ gives
\begin{equation}
 \frac{\dd}{\dd r_0}
 \left(\frac{h_{n,0}}{k_0}\right)
 =(n+1)\frac{h_{n+1,0}}{k_0},
 \qquad 1\le n\le N-1.
 \label{eq:normalized_h_recurrence}
\end{equation}

At the quartic orbit, $h_{1,m}=h_{2,m}=h_{3,m}=0$.
We use this recurrence to expand $h_{n,0}/k_0$ about $r_0=r_m$.
Substituting these expansions into Eq.~\eqref{eq:Taylor5} and collecting terms gives
\begin{equation}
\begin{aligned}
 \frac{F(r,r_0)}{k_0}
 ={}&h_{4,m}Q_{4,m}+h_{5,m}Q_{5,m}\\
 &+O\!\left((r-r_0)(r-r_m)^5\right),\\
 Q_{n,m}(r,r_0)
 ={}&(r-r_m)^n-(r_0-r_m)^n,\quad n=4,5.
\end{aligned}
\label{eq:quartic_F_polynomial}
\end{equation}

To expand $[F(r,r_0)/k_0]^{-1/2}$, we use the following relations for $r>r_0$:
\begin{equation}
\begin{aligned}
 Q_{5,m}
 &=(r-r_m)Q_{4,m}
   +(r_0-r_m)^4(r-r_0),\\
 Q_{4,m}
 &\ge (r-r_0)(r-r_m)^3,\\
 0<\frac{Q_{5,m}}{Q_{4,m}}
 &\le\frac54(r-r_m).
\end{aligned}
\label{eq:Q5identity}
\end{equation}
Near $r_m$, Eqs.~\eqref{eq:Q5identity} and~\eqref{eq:prefactor_local} give
\begin{equation}
\begin{aligned}
 \frac{\bar S(r,r_0)}{\sqrt{F(r,r_0)}}
 &=f_4(r,r_0)+\mathcal E_4,\\
 f_4(r,r_0)
 &=
 \frac{S_m}{\sqrt{h_{4,m}}\sqrt{Q_{4,m}}}
 +\frac{T_{r,m}}{\sqrt{h_{4,m}}}
 \frac{r-r_m}{\sqrt{Q_{4,m}}}\\
 &\quad-
 \frac{S_mh_{5,m}}{2h_{4,m}^{3/2}}
 \frac{(r_0-r_m)^4(r-r_0)}
 {Q_{4,m}^{3/2}},\\
 |\mathcal E_4|
 &\le K\frac{(r-r_m)^2}{\sqrt{Q_{4,m}}}.
\end{aligned}
\label{eq:cusp_integrand_expansion}
\end{equation}
The bound holds with the same $K>0$ for all $r_0$ sufficiently close to $r_m$.
The coefficient $T_{r,m}$ is defined in Eq.~\eqref{eq:quartic_critical_density}.

The deflection integral in Eq.~\eqref{eq:original} converges at infinity.
At fixed $r_0$, the definition in Eq.~\eqref{eq:quartic_F_polynomial} gives $Q_{4,m}\sim r^4$ as $r\to\infty$.
The first term in $f_4$ is $O(r^{-2})$.
The third is $O(r^{-5})$.
Both are integrable at infinity.
In the second term, $(r-r_m)/\sqrt{Q_{4,m}}\sim1/r$.
This factor is not integrable at infinity.

To remove this $1/r$ tail, we define the auxiliary integral
\begin{equation}
 I_A(r_0)=
 \int_{r_0}^{\infty}
 \left[
 f_4(r,r_0)
 -\frac{T_{r,m}}
 {\sqrt{h_{4,m}}(r-r_m+R_*)}
 \right]\dd r.
 \label{eq:quartic_auxiliary_integral}
\end{equation}
Here $R_*>0$ is a constant.
At $r=r_m$, the subtracted function $1/(r-r_m+R_*)$ has the finite value $1/R_*$.
The subtraction therefore leaves the $(r-r_m)^{-1}$ term in Eq.~\eqref{eq:quartic_critical_density} unchanged.

The remaining integral is
\begin{equation}
 I_R(r_0)
 =I(r_0)-I_A(r_0)
 =\int_{r_0}^{\infty}g_4(r,r_0)\,\dd r.
\label{eq:quartic_integral_split}
\end{equation}
Here $g_4(r,r_0)$ is the integrand in Eq.~\eqref{eq:original} minus the integrand of $I_A$.

We first evaluate $I_A$.
To evaluate the first term in $f_4$, set $z=(r-r_m)/(r_0-r_m)$.
The required integral is
\begin{equation}
\begin{aligned}
 \int_{r_0}^{\infty}
 \frac{\dd r}{\sqrt{Q_{4,m}}}
 &=\frac{J_4}{r_0-r_m},\\
 J_4
 &=\int_1^{\infty}\frac{\dd z}{\sqrt{z^4-1}}
 =\frac14 B\!\left(\frac14,\frac12\right).
\end{aligned}
\label{eq:J4}
\end{equation}
Here $J_4\simeq1.31103$.
This integral gives the $(r_0-r_m)^{-1}$ divergence.

For the second term and the subtraction in $I_A$, we use
\begin{equation}
\begin{aligned}
 &\int_{r_0}^{\infty}
 \left[
 \frac{r-r_m}{\sqrt{Q_{4,m}}}
 -\frac{1}{r-r_m+R_*}
 \right]\dd r\\
 &=\ln\frac{r_0-r_m+R_*}{r_0-r_m}
  +\frac12\ln2.
\end{aligned}
\label{eq:quartic_aux_limits}
\end{equation}
This integral diverges logarithmically as $r_0\to r_m^+$.
It gives a logarithmic term in $\Theta$ when $T_{r,m}\ne0$.

For the third term in $f_4$, the required integral is
\begin{equation}
\begin{aligned}
 K_4
 &=\int_{r_0}^{\infty}
 \frac{(r_0-r_m)^4(r-r_0)}
 {Q_{4,m}^{3/2}}\,\dd r\\
 &=\int_1^{\infty}
 \frac{z-1}{(z^4-1)^{3/2}}\,\dd z
 =\frac{J_4-1}{2}.
\end{aligned}
\label{eq:K4finite}
\end{equation}
The third term therefore contributes the finite quantity $-S_mh_{5,m}K_4/(2h_{4,m}^{3/2})$.

We now determine the limit of $I_R(r_0)$ in Eq.~\eqref{eq:quartic_integral_split}.
We first justify taking $r_0\to r_m^+$ inside the integral.

Take $\eta>0$ sufficiently small and $0<r_0-r_m<\eta$.
For $r_0<r<r_m+\eta$, the expansion and bounds in Eqs.~\eqref{eq:cusp_integrand_expansion} and~\eqref{eq:Q5identity} give
\begin{gather}
 g_4(r,r_0)
 =\mathcal E_4+
 \frac{T_{r,m}}
 {\sqrt{h_{4,m}}(r-r_m+R_*)},
 \label{eq:quartic_remainder}\\[1mm]
 \int_{r_0}^{r_m+\eta}
 |g_4(r,r_0)|\,\dd r
 \le
 \left(
 3K+\frac{|T_{r,m}|}{\sqrt{h_{4,m}}R_*}
 \right)\eta.
 \label{eq:quartic_residual_bound}
\end{gather}
This contribution tends to zero as $\eta\to0^+$, uniformly for $0<r_0-r_m<\eta$.

Away from $r_m$, the integrands in Eqs.~\eqref{eq:original} and~\eqref{eq:quartic_auxiliary_integral} are continuous in $(r,r_0)$ up to $r_0=r_m$.
Their difference $g_4$ therefore converges uniformly on fixed finite intervals.

At infinity, both integrands are $O(r^{-2})$ uniformly for $r_0$ near $r_m$.
The integral of $|g_4|$ over sufficiently large $r$ can therefore be made uniformly small.

These estimates allow us to take the limit $r_0\to r_m^+$ inside $I_R$.
At fixed $r>r_m$, $Q_{4,m}\to(r-r_m)^4$ and the last term in $f_4$ tends to zero.
We therefore obtain
\begin{equation}
\begin{gathered}
 I_R(r_0)=\mathcal H_4+o(1),\\[1mm]
 \begin{aligned}
 \mathcal H_4
 &=\int_{r_m}^{\infty}\Bigg[
 \frac{\bar S(r,r_m)}{\sqrt{F(r,r_m)}}
 -\frac{S_m}{\sqrt{h_{4,m}}(r-r_m)^2}\\
 &\qquad
 -\frac{T_{r,m}}{\sqrt{h_{4,m}}}
 \left(
 \frac1{r-r_m}
 -\frac1{r-r_m+R_*}
 \right)
 \Bigg]\,\dd r.
 \end{aligned}
\end{gathered}
\label{eq:Bfinite}
\end{equation}

Using $\Theta=I_A+I_R-\pi$, we obtain the expansion as $r_0\to r_m^+$:
\begin{equation}
\begin{aligned}
 \Theta(r_0)&=
 \frac{S_mJ_4}
 {\sqrt{h_{4,m}}(r_0-r_m)}\\
 &+\frac{T_{r,m}}{\sqrt{h_{4,m}}}
 \left[
 \ln\frac{r_0-r_m+R_*}{r_0-r_m}
 +\frac12\ln2
 \right]\\
 &-\frac{S_mh_{5,m}}{2h_{4,m}^{3/2}}K_4
 +\mathcal H_4-\pi+o(1).
\end{aligned}
\label{eq:quartic_angle_r0}
\end{equation}

We now express Eq.~\eqref{eq:quartic_angle_r0} in terms of $\eps=b/b_m-1$.
Equations~\eqref{eq:exact_split} and~\eqref{eq:quartic_critical_root} give
\begin{gather}
 \eps
 =\frac{h_{4,m}(r_0-r_m)^4+h_{5,m}(r_0-r_m)^5}{2A_m}
 +O((r_0-r_m)^6),
 \notag\\*
 \frac{1}{r_0-r_m}
 =\left(\frac{h_{4,m}}{2A_m}\right)^{1/4}\eps^{-1/4}
 +\frac{h_{5,m}}{4h_{4,m}}+o(1),
 \label{eq:cusp_coefficient}\\*
 \ln\frac{r_0-r_m+R_*}{r_0-r_m}
 =\frac14\ln\frac{h_{4,m}R_*^4}{2A_m\eps}+o(1).
 \notag
\end{gather}

Substituting Eq.~\eqref{eq:cusp_coefficient} into Eq.~\eqref{eq:quartic_angle_r0} and simplifying the constant terms using Eq.~\eqref{eq:K4finite}, we obtain
\begin{align}
 \Theta
 &=\bar c\,\eps^{-1/4}+\ell\ln\eps+d+o(1),
 \quad \eps\to0^+,
 \label{eq:cusp_law}\\
 \bar c
 &=\frac{S_mJ_4}
 {h_{4,m}^{1/4}(2A_m)^{1/4}},
 \qquad
 \ell=-\frac{T_{r,m}}{4\sqrt{h_{4,m}}},
 \label{eq:cusp_log_coefficient}\\
 d
 &=\mathcal H_4-\pi
 +\frac{S_mh_{5,m}}{4h_{4,m}^{3/2}}
 -\ell\ln\frac{2h_{4,m}R_*^4}{A_m}.
 \label{eq:finite4}
\end{align}
The finite term $d$ is independent of $R_*$.

Both $\bar c$ and $\ell$ depend on the geometry near the orbit, whereas $d$ depends on the full returning path.
The coefficient $\ell$ involves $h_{5,m}$ and $S_{1,m}$, which do not enter $\bar c$.
It therefore provides a further constraint on the local geometry beyond the leading coefficient.

We next apply the expansion to specific geometries and show that $\ell$ can vary while the critical impact parameter and $\bar c$ remain fixed.

\section{Deflection angles in wormhole and black-hole spacetimes}
\label{sec:geometry}

The critical impact parameter and the leading deflection coefficient do not fully determine the strong-deflection behavior near a quartic photon orbit.
We examine the additional geometric information carried by the logarithmic term in the Simpson--Visser wormhole, the hairy Schwarzschild black hole, and a black hole with a compact Hernquist-type matter halo.

The subscript $c$ denotes the quartic orbit, corresponding to $m$ in the general formulas.
We write its critical impact parameter as $b_c$ and use $\eps=b/b_c-1>0$.

Table~\ref{tab:quartic_coefficients} lists $b_c$, $\bar c$, $\ell$, and $d$ for the three configurations.
The deflection coefficients follow from Eqs.~\eqref{eq:cusp_log_coefficient} and~\eqref{eq:finite4}, with $\mathcal H_4$ evaluated numerically.

We integrate Eq.~\eqref{eq:original} numerically to obtain $\Theta_{\rm num}$.
The turning-point offsets are $10^{-k}$, $k=1,\ldots,8$, in the mass units used for each model.

\begin{figure*}[t]
\centering
\includegraphics[width=0.9\textwidth]{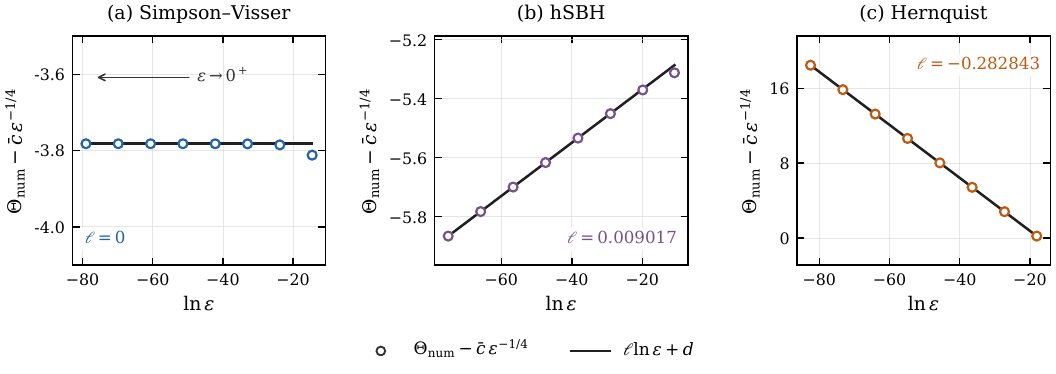}
\caption{Subleading deflection behavior of the Simpson--Visser (a), hSBH (b), and Hernquist (c) configurations.
Open circles give $\Theta_{\rm num}-\bar c\eps^{-1/4}$, and solid black lines give $\ell\ln\eps+d$.
Their vertical separation gives the absolute error of the three-term expansion.
The limit $\eps\to0^+$ is toward the left.
The panels use different vertical scales.}
\label{fig:deflection_residuals}
\end{figure*}

To examine the logarithmic contribution separately from the leading divergence, we subtract the leading power-law term from the deflection angle.
Equation~\eqref{eq:cusp_law} gives
\begin{equation}
 \Theta-\bar c\eps^{-1/4}
 =\ell\ln\eps+d+o(1),
 \qquad \eps\to0^+.
 \label{eq:geometry_residual}
\end{equation}
Figure~\ref{fig:deflection_residuals} compares $\Theta_{\rm num}-\bar c\eps^{-1/4}$ with the theoretical lines $\ell\ln\eps+d$ as functions of $\ln\eps$.
The vertical distance between each numerical point and the corresponding theoretical line equals the absolute difference between $\Theta_{\rm num}$ and the approximation $\bar c\eps^{-1/4}+\ell\ln\eps+d$.
Table~\ref{tab:quartic_convergence} reports these errors for $k=2,4,6$.

To check the logarithmic coefficient, we fit $\Theta_{\rm num}-\bar c\eps^{-1/4}$ as a linear function of $\ln\eps$ using the four numerical samples with $k=5,\ldots,8$.
Both the slope and intercept are varied in the least-squares fit.

\begin{table}[t]
\caption{Critical impact parameters and strong-deflection coefficients for the Simpson--Visser, hSBH, and Hernquist configurations.
The units of $b_c$ are $m$ for Simpson--Visser, $M$ for hSBH, and $M_{\rm BH}$ for Hernquist.
The same values of $d$ are obtained from Eq.~\eqref{eq:finite4} with $R_*=1$ and $2.5$ in the corresponding mass units.}
\label{tab:quartic_coefficients}
\centering
\setlength{\tabcolsep}{3pt}
\begin{ruledtabular}
\begin{tabular}{lrrr}
& Simpson--Visser & hSBH & Hernquist \\
$b_c$ & $5.1961524227$ & $4.6026581073$ & $28.9582702622$ \\
$\bar c$ & $4.1037418463$ & $4.7078236869$ & $4.8801990017$ \\
$\ell$ & $0$ & $0.0090171434$ & $-0.2828427125$ \\
$d$ & $-3.7819207290$ & $-5.1884899627$ & $-4.8903304871$ \\
\end{tabular}
\end{ruledtabular}
\end{table}

\begin{table}[t]
\caption{Absolute errors of $\bar c\eps^{-1/4}+\ell\ln\eps+d$ compared with the numerical deflection angle $\Theta_{\rm num}$.
The errors are in radians.
The turning-point offsets are $10^{-k}$ in each model's mass units.
The errors are calculated using unrounded numerical results and coefficients.}
\label{tab:quartic_convergence}
\centering
\setlength{\tabcolsep}{3pt}
\begin{ruledtabular}
\begin{tabular}{lcrr}
Configuration & $k$ & $\eps$ & Absolute error \\
Simpson--Visser & 2 & $4.630\times10^{-11}$ & $2.995\times10^{-3}$ \\
& 4 & $4.630\times10^{-19}$ & $2.995\times10^{-5}$ \\
& 6 & $4.630\times10^{-27}$ & $2.995\times10^{-7}$ \\
\addlinespace[2pt]
hSBH & 2 & $2.360\times10^{-9}$ & $2.776\times10^{-3}$ \\
& 4 & $2.418\times10^{-17}$ & $2.794\times10^{-5}$ \\
& 6 & $2.419\times10^{-25}$ & $2.794\times10^{-7}$ \\
\addlinespace[2pt]
Hernquist & 2 & $1.441\times10^{-12}$ & $5.925\times10^{-4}$ \\
& 4 & $1.447\times10^{-20}$ & $5.930\times10^{-6}$ \\
& 6 & $1.447\times10^{-28}$ & $5.930\times10^{-8}$ \\
\end{tabular}
\end{ruledtabular}
\end{table}

\subsection{Simpson--Visser wormhole}
\label{sec:symmetric_example}

We first consider the Simpson--Visser wormhole \cite{SimpsonVisser2019,Tsukamoto2021SimpsonVisser}.
We use a regular radial coordinate $\rho$, with the throat at $\rho=0$.
The metric functions are
\begin{equation}
 R=\sqrt{\rho^2+a_{\rm SV}^2},
 \quad A=1-\frac{2m}{R},\quad
 B=A^{-1},\quad C=R^2.
\label{eq:sv_metric}
\end{equation}
Here $m>0$, and the wormhole requires $a_{\rm SV}>2m$.

Equations~\eqref{eq:effective_potential} and~\eqref{eq:sv_metric} give
\begin{equation}
 \frac{\dd V}{\dd\rho}
 =-\frac{2\rho(R-3m)}{R^5}.
\end{equation}
For $2m<a_{\rm SV}<3m$, two unstable circular null orbits at $\rho=\pm\sqrt{9m^2-a_{\rm SV}^2}$ are separated by a stable orbit at the throat.
All three merge at $\rho_c=0$ when $a_{\rm SV}=3m$.

At this parameter value,
\begin{equation}
 V(\rho)=\frac1{27m^2}
 -\frac{\rho^4}{2916m^6}+O(\rho^6).
\label{eq:sv_local}
\end{equation}
Thus $V$ has a quartic maximum at the throat, where $A_c=1/3$, $B_c=3$, $C_c=9m^2$, and $b_c=3\sqrt3m$.

For $\rho>0$, $V$ decreases monotonically to zero.
A ray arriving from $\rho\to+\infty$ with $b>b_c$ therefore turns at $\rho_0>0$ and returns to the same asymptotic end.

At $\rho_0=\rho_c=0$, the numerator in Eq.~\eqref{eq:original} is $\bar S(\rho,0)=6m/(\rho^2+9m^2)$.
Both $V(\rho)$ and $\bar S(\rho,0)$ are even in $\rho$, so $h_{5,c}=S_{1,c}=0$.
The other local coefficients are $h_{4,c}=1/(324m^4)$ and $S_c=2/(3m)$.
Equation~\eqref{eq:cusp_log_coefficient} then gives
\begin{equation}
 \bar c=2^{5/4}3^{1/4}J_4,\qquad \ell=0.
\label{eq:sv_coefficients}
\end{equation}
The leading coefficient agrees with the numerical estimate in Ref.~\cite{Tsukamoto2026Z2Throat}.

With $\ell=0$, Eq.~\eqref{eq:geometry_residual} predicts $\Theta-\bar c\eps^{-1/4}\to d$ as $\eps\to0^+$.
The numerical points in Fig.~\ref{fig:deflection_residuals}(a) approach this horizontal line.
The linear fit to the numerical points gives a slope of $-1.00\times10^{-7}$, close to zero.

\subsection{Hairy Schwarzschild black hole}
\label{sec:nonzero_examples}

We next consider the hairy Schwarzschild black hole obtained by gravitational decoupling~\cite{Ovalle2021}.
We use $x=r/M$ and $\tau=t/M$, where $M>0$ is the asymptotic mass.
For the dimensionless line element $\dd s^2/M^2$, the metric functions are
\begin{equation}
\begin{aligned}
 A&=f(x),\qquad B=f(x)^{-1},\qquad C=x^2,\\
 f(x)&=1-\frac2x+\alpha\ee^{-x/\mu},
 \qquad \mu=1-\lambda.
\end{aligned}
\label{eq:metric}
\end{equation}
Here $0<\lambda<1$ and $\alpha\ge0$.
The parameters $\alpha$ and $\mu$ control the amplitude and decay length of the deformation, respectively.
The Schwarzschild limit is $\alpha=0$.

Previous work has calculated the logarithmic strong-deflection coefficients and the corresponding lensing observables~\cite{Jha2023}.
Multiple photon spheres, accretion images, and interferometric signatures have also been studied~\cite{Guo2023Multiple,Meng2023Images, Wang2025Interferometric}.
We measure $b$ in units of $M$.
Equation~\eqref{eq:effective_potential} gives $V=f/x^2$.
Primes in this subsection denote derivatives with respect to $x$.

For suitable parameters, $V$ has two maxima separated by a minimum.
The two unstable orbits and the stable orbit between them merge at
\begin{equation}
\begin{aligned}
 \lambda_c&=16-9\sqrt3,\qquad x_c=12-6\sqrt3,\\
 \alpha_c&=\frac{\sqrt3-1}{2}\ee^{1+\sqrt3}.
\end{aligned}
\label{eq:Cgeometry}
\end{equation}
At this orbit, $V'_c=V''_c=V'''_c=0$ and $V^{(4)}_c<0$, giving a quartic maximum.
The metric function satisfies $f_c=(\sqrt3-1)/6>0$.

Numerically, the outermost horizon is located at $x_h\simeq0.909705$, inside the quartic orbit at $x_c\simeq1.607695$.
For $x>x_c$, $V$ decreases monotonically to zero.
Rays from infinity with $b>b_c=x_c/\sqrt{f_c}$ therefore turn at a unique $x_0>x_c$ and return to infinity.

Equation~\eqref{eq:original} gives $\bar S(x,x_0)=2x_0/x^2$.
Using Eqs.~\eqref{eq:Taylor5} and~\eqref{eq:prefactor_local}, we obtain the local coefficients
\begin{equation}
\begin{aligned}
 S_c&=\frac2{x_c},\qquad S_{1,c}=-\frac4{x_c^2},\\
 h_{4,c}&=\frac{3+\sqrt3}{12x_c^4},\qquad
 \frac{h_{5,c}}{h_{4,c}}
 =-\frac{51+5\sqrt3}{15x_c}.
\end{aligned}
\label{eq:h01}
\end{equation}
Substitution into Eq.~\eqref{eq:cusp_log_coefficient} gives
\begin{equation}
 \bar c=(96\sqrt3)^{1/4}J_4,\qquad
 \ell=\frac{\sqrt3-1}{10\sqrt{33+19\sqrt3}}>0.
\label{eq:cusp_exact_log}
\end{equation}

Since $\ell>0$, Eq.~\eqref{eq:geometry_residual} predicts $\Theta-\bar c\eps^{-1/4}\to-\infty$ as $\eps\to0^+$.
The numerical points in Fig.~\ref{fig:deflection_residuals}(b) approach the theoretical line $\ell\ln\eps+d$.
The linear fit to the numerical points gives a slope of $0.00901705$, compared with the theoretical value $\ell\simeq0.00901714$.

\subsection{Black hole in a Hernquist-type matter halo}
\label{sec:hernquist_example}

We next consider a black hole surrounded by a Hernquist-type matter halo~\cite{Cardoso2022Halo}.
Let $M_{\rm BH}$ be the central black-hole mass parameter, and let $M_h$ and $a_0$ be the halo mass and radial scale.
We use $r=R/M_{\rm BH}$, $\tau=t/M_{\rm BH}$, $a=a_0/M_{\rm BH}$, and $H=M_h/M_{\rm BH}$, with $b$ measured in units of $M_{\rm BH}$.
For the dimensionless line element $\dd s^2/M_{\rm BH}^2$, the metric functions are
\begin{equation}
\begin{aligned}
 A&=\left(1-\frac2r\right)\ee^{\Upsilon},
 \qquad C=r^2,\\
 B&=\left[1-\frac{2\mathfrak m(r)}r\right]^{-1},
 \qquad \mathfrak m(r)=1+H\frac{(r-2)^2}{(r+a)^2},\\
 \Upsilon&=-\pi\sqrt{\frac H\xi}
 +2\sqrt{\frac H\xi}\arctan\frac{r+a-H}{\sqrt{H\xi}},\\
 \xi&=2a-H+4.
\end{aligned}
\label{eq:hernquist_metric}
\end{equation}
We take $a>0$ and $0<H<2a+4$, so that $\xi>0$.
The radial metric function can be written as
\begin{equation}
\begin{aligned}
 B^{-1}&=\frac{(r-2)D(r)}{r(r+a)^2},\\
 D(r)&=(r+a-H)^2+H\xi>0.
\end{aligned}
\end{equation}
Both $A$ and $B^{-1}$ are positive for $r>2$ and vanish at the horizon $r=2$.

A sufficiently compact halo can support multiple photon spheres~\cite{Xavier2023Halo}.
For $V=A/r^2$, we find
\begin{equation}
\begin{aligned}
 \frac{V'}V&=-\frac{2P(r)}{r(r-2)D(r)},\\
 P(r)&=(r-3)(r+a)^2-3H(r-2)^2.
\end{aligned}
\label{eq:hernquist_polynomials}
\end{equation}
The exterior roots of $P$ determine the circular null orbits.
For suitable halo parameters, two unstable orbits are separated by a stable orbit outside the horizon.
Their merger requires a triple root of $P$.
Solving $P=P'=P''=0$ gives $(r_c,a,H)=(6,6,9)$ in the parameter range above.

At these parameters, $P=(r-6)^3$, and
\begin{equation}
 \frac{V'}V=-\frac{2(r-6)^3}
 {r(r-2)[(r-3)^2+63]}.
\label{eq:hernquist_selected_path}
\end{equation}
Thus $V$ has a quartic maximum at $r_c=6$, outside the horizon at $r=2$.
For $r>6$, $V$ decreases monotonically to zero.
Rays from infinity with $b>b_c=V(6)^{-1/2}$ therefore turn at $r_0>6$ and return to infinity.

At $r_0=r_c=6$, the numerator in Eq.~\eqref{eq:original} is $\bar S(r,r_c)=12\sqrt{A(r)B(r)}/r^2$.
With $A_c=A(6)$, $B_c=3$, and $C_c=36$, the local coefficients are
\begin{equation}
 h_{4,c}=\frac{A_c}{3456},
 \quad \frac{h_{5,c}}{h_{4,c}}=-\frac25,\quad
 S_c=\sqrt{\frac{A_c}{3}},
 \quad \frac{S_{1,c}}{S_c}=-\frac16.
\end{equation}
Equation~\eqref{eq:cusp_log_coefficient} then gives
\begin{equation}
 \bar c=192^{1/4}J_4,\qquad
 \ell=-\frac{\sqrt2}{5}.
\label{eq:hernquist_coefficients}
\end{equation}

Since $\ell<0$, Eq.~\eqref{eq:geometry_residual} predicts $\Theta-\bar c\eps^{-1/4}\to+\infty$ as $\eps\to0^+$.
The numerical points in Fig.~\ref{fig:deflection_residuals}(c) approach the theoretical line $\ell\ln\eps+d$.
The linear fit to the numerical points gives a slope of $-0.28284273$, compared with the theoretical value $\ell\simeq-0.28284271$.

For all three geometries, Table~\ref{tab:quartic_convergence} shows that the absolute difference between $\Theta_{\rm num}$ and $\bar c\eps^{-1/4}+\ell\ln\eps+d$ decreases by about four orders of magnitude from $k=2$ to $k=6$.
At $k=6$, all three errors are below $3.0\times10^{-7}$ rad.
These comparisons confirm the strong-deflection expansion in all three geometries.

\section{Geometric information in high-order images}
\label{sec:images}

We now relate the leading and logarithmic deflection coefficients to high-order image positions.
We show how these positions can be used to infer $\bar c$ and $\ell$ and constrain the geometry near the quartic orbit.

We consider the returning rays of Sec.~\ref{sec:returning_general}.
A point source and a static observer are fixed in the same asymptotic region at $r_S$ and $r_O$, respectively, with an oriented azimuthal separation $0\leq\Delta\phi<2\pi$.
The quartic orbit is at $r=r_c$, with critical impact parameter $b_c$.
Each ray turns at $r_c<r_0<\min(r_S,r_O)$.
For a fixed sign of angular momentum, the azimuthal change is $2\pi n+\Delta\phi$, where $n$ counts the additional full turns and labels the images in the chosen sequence.

Equation~\eqref{eq:original} gives $I=\Theta+\pi$ for a ray with both endpoints at infinity.
For finite $r_S$ and $r_O$, we subtract the azimuthal changes from these radii to infinity.
This gives the image equation~\cite{Perlick2004, BozzaScarpetta2007,Bozza2008LensEquation}
\begin{equation}
\begin{aligned}
 \Theta(b_n)={}&2\pi n+\Delta\phi-\pi\\
 &+\sum_{j=S,O}\int_{r_j}^{\infty}
 \frac{\sqrt{A(r)B(r)}}
 {C(r)\sqrt{b_n^{-2}-V(r)}}\,\dd r.
\end{aligned}
\label{eq:lens_winding}
\end{equation}

Let $\delta_c$ be $\Delta\phi-\pi$ plus the two finite-distance integrals evaluated at $b=b_c$.
Their smooth dependence on $b$ gives
\begin{equation}
 \Theta(b_n)=2\pi n+\delta_c+O(\eps_n),
 \qquad \eps_n=b_n/b_c-1\to0^+.
 \label{eq:image_geometry_c}
\end{equation}

We measure the apparent angle $\vartheta_n$ from the inward radial direction at the observer.
Let $\vartheta_c$ be its value at $b=b_c$.
The local photon momentum gives~\cite{Perlick2004, PerlickTsupko2022}
\begin{equation}
\begin{gathered}
 \sin\vartheta_n=b_n\sqrt{\frac{A(r_O)}{C(r_O)}},
 \quad
 \sin\vartheta_c=b_c\sqrt{\frac{A(r_O)}{C(r_O)}},\\
 \vartheta_n-\vartheta_c
 =\tan\vartheta_c\,\eps_n+O(\eps_n^2).
\end{gathered}
\label{eq:image_observer_angle}
\end{equation}
We take $0<\vartheta_c<\vartheta_n<\pi/2$.
The images approach the critical circle of angular radius $\vartheta_c$ from outside.

The leading balance in Eqs.~\eqref{eq:cusp_law} and~\eqref{eq:image_geometry_c} gives $\bar c\eps_n^{-1/4}\sim2\pi n$, and hence $\eps_n\sim[\bar c/(2\pi n)]^4$.
Keeping the next terms and using Eq.~\eqref{eq:image_observer_angle}, we obtain
\begin{equation}
\begin{aligned}
 \vartheta_n-\vartheta_c
 &=\frac{\bar c^4\tan\vartheta_c}{(2\pi n)^4}
 -\frac{16\ell\bar c^4\tan\vartheta_c}{(2\pi n)^5}
 \ln\frac{2\pi n}{\bar c}\\
 &\quad+\frac{4\bar c^4\tan\vartheta_c}{(2\pi n)^5}
 (d-\delta_c)+o(n^{-5}).
\end{aligned}
\label{eq:cusp_image_correction}
\end{equation}
The image offsets decrease as $n^{-4}$.
The logarithmic deflection term contributes at order $n^{-5}\ln n$.
The nonlogarithmic correction has the smaller asymptotic order $n^{-5}$ and includes the effects of $d$ and $\delta_c$.
The $b$-dependence of the finite-distance integrals and the nonlinear angular terms contribute only beyond order $n^{-5}$.

Let $q_0$ and $q_1$ be the coefficients of $n^{-4}$ and $n^{-5}\ln n$ in Eq.~\eqref{eq:cusp_image_correction}.
Then
\begin{equation}
\begin{aligned}
 q_0&=\frac{\bar c^4\tan\vartheta_c}{(2\pi)^4},\qquad
 q_1=-\frac{16\ell\bar c^4\tan\vartheta_c}{(2\pi)^5},\\
 \bar c&=2\pi
 \left(\frac{q_0}{\tan\vartheta_c}\right)^{1/4},
 \qquad
 \ell=-\frac{\pi}{8}\frac{q_1}{q_0}.
\end{aligned}
\label{eq:image_geometric_ratio}
\end{equation}
The critical angle $\vartheta_c$ and the leading image coefficient $q_0$ determine $\bar c$.
The ratio $q_1/q_0$ cancels the common factor $\bar c^4\tan\vartheta_c$ and gives $\ell$.
This ratio is independent of the source and observer positions and does not contain $d$ or $\delta_c$.

Appendix~\ref{app:halo_deformation} constructs a family of halo geometries with the same $b_c$ and $\bar c$ but different $\ell$.
With the source and observer fixed at the same positions outside the modified region, these geometries have the same $\vartheta_c$ and $q_0$ but different $q_1$.
The logarithmic correction therefore distinguishes geometries whose critical circles and leading image offsets agree.

To display the correction to the leading image law, we take the inverse fourth root of the angular offsets.
For the leading term alone, $[q_0/(\vartheta_n-\vartheta_c)]^{1/4}=n$.
We therefore define
\begin{equation}
 P_n=
 \left(
 \frac{q_0}{\vartheta_n-\vartheta_c}
 \right)^{1/4}-n.
 \label{eq:image_scaled_offset}
\end{equation}
Equation~\eqref{eq:cusp_image_correction} gives
\begin{equation}
 P_n=\frac{2\ell}{\pi}\ln\frac{2\pi n}{\bar c}
 +\frac{\delta_c-d}{2\pi}+o(1),
 \qquad n\to\infty.
\label{eq:image_scaled_law}
\end{equation}
As $n$ increases, $P_n$ approaches a straight line in $\ln n$, with slope $2\ell/\pi$.
The finite term $d$ and the source--observer contribution $\delta_c$ enter only the constant part.

For the three configurations in Table~\ref{tab:quartic_coefficients}, we take $r_S=r_O=1000$ in each model's mass units and $\Delta\phi=\pi-0.1$.
For each $n$, we solve Eq.~\eqref{eq:lens_winding} for $b_n$ using the numerical deflection angle from Eq.~\eqref{eq:original}.
The finite-distance integrals are evaluated at $b_n$.
The sine relation in Eq.~\eqref{eq:image_observer_angle} then gives the numerical image positions, from which we calculate $P_n$.

\begin{figure*}[t]
\centering
\includegraphics[width=0.9\textwidth]{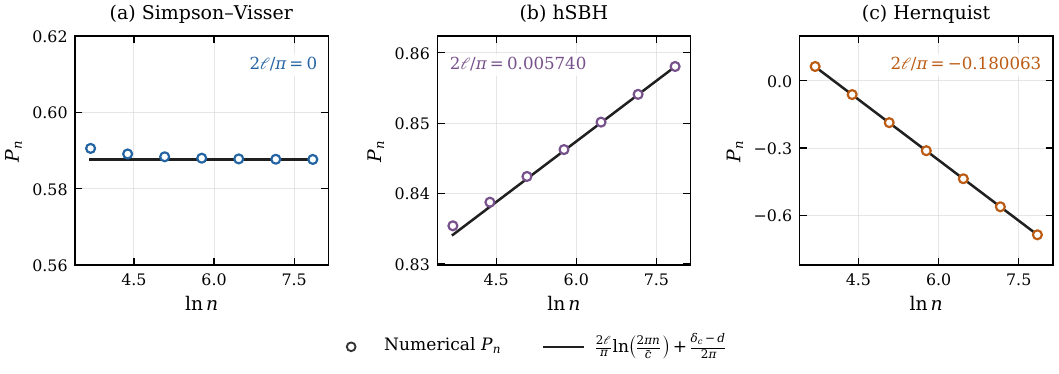}
\caption{Subleading structure of high-order images in the Simpson--Visser (a), hSBH (b), and Hernquist (c) configurations.
Open circles give $P_n$ calculated from numerical image positions using Eq.~\eqref{eq:image_scaled_offset}.
Solid black lines give $(2\ell/\pi)\ln(2\pi n/\bar c)+(\delta_c-d)/(2\pi)$.
The source and observer radial coordinates are $1000$ in each model's mass units, with $\Delta\phi=\pi-0.1$.
The image orders are $n=40,80,160,320,640,1280,2560$ in all three panels.}
\label{fig:image_accumulation}
\end{figure*}

Figure~\ref{fig:image_accumulation} compares the numerical $P_n$ with the lines predicted by Eq.~\eqref{eq:image_scaled_law}.
The lines use the coefficients in Table~\ref{tab:quartic_coefficients} and $\delta_c$.

In Fig.~\ref{fig:image_accumulation}(a), the Simpson--Visser points approach the predicted constant, consistent with $\ell=0$.
The hSBH and Hernquist points approach the predicted lines with slopes $2\ell/\pi\simeq0.00574$ and $-0.18006$ in panels (b) and (c), respectively.
At sufficiently large $n$, the logarithmic correction reduces the hSBH image offsets and increases the Hernquist offsets relative to the leading $n^{-4}$ prediction in each geometry.

For a resolved high-order image sequence, the angular positions can be fitted for $\vartheta_c$, $q_0$, $q_1$, and the coefficient $q_2$ of the nonlogarithmic $n^{-5}$ term.
If the relative image orders are known but the absolute starting order is not, we write $n=j+s$, where $j$ is the relative label and $s$ is a fixed integer.
Equation~\eqref{eq:cusp_image_correction} then gives
\begin{equation}
 \vartheta_{j+s}-\vartheta_c
 =q_0j^{-4}+q_1j^{-5}\ln j
 +(q_2-4sq_0)j^{-5}+o(j^{-5}).
\label{eq:image_order_shift}
\end{equation}
The coefficients $q_0$ and $q_1$ are unchanged, while the nonlogarithmic coefficient becomes $q_2-4sq_0$.
Thus the asymptotic relations for $\bar c$ and $\ell$ in Eq.~\eqref{eq:image_geometric_ratio} do not depend on a fixed shift of the starting image order.

These two coefficients constrain the geometry near the quartic orbit.
The halo construction shows that the constraint from $\ell$ can distinguish geometries with the same critical circle and leading image coefficient.
Candidate spacetime metrics can therefore be tested against both the leading image accumulation and its logarithmic correction.

\section{Conclusions}
\label{sec:conclusion}

We have shown that the deflection angle can contain both power-law and logarithmic divergences near a quartic maximum of the null effective potential in asymptotically flat, static, and spherically symmetric spacetimes.
For rays that turn outside the quartic orbit and return to the same asymptotic region, we obtain
\[
 \Theta=\bar c\eps^{-1/4}+\ell\ln\eps+d+o(1),
 \qquad \eps=b/b_m-1\to0^+.
\]
Both $\bar c$ and $\ell$ are determined by the geometry near the orbit.
The finite term $d$ depends on the geometry along the full returning path.

Reflection symmetry gives $\ell=0$ at the Simpson--Visser throat considered here.
The hSBH and Hernquist examples have $\ell>0$ and $\ell<0$, respectively.
Numerical deflection angles confirm the expansion in these examples.

For fixed source and observer positions, the leading image offsets from the critical circle decrease as $n^{-4}$.
The logarithmic deflection coefficient $\ell$ controls the $n^{-5}\ln n$ correction to the image offsets.
Numerical image sequences confirm the predicted asymptotic structure in these examples.

The leading image coefficient $q_0$ and the critical angle $\vartheta_c$ determine $\bar c$.
The ratio of the logarithmic image coefficient $q_1$ to $q_0$ gives
\[
 \ell=-\frac{\pi}{8}\frac{q_1}{q_0}.
\]
This ratio is independent of the source and observer positions and does not contain the finite term $d$.
By locally modifying the halo geometry, we also construct geometries with the same critical impact parameter and $\bar c$ but different $\ell$.
For the locally modified halo geometries, placing the source and observer at the same fixed positions beyond the modified region gives the same critical angle and leading image coefficient, but different logarithmic corrections.

Our results provide an additional geometric constraint for testing candidate spacetime metrics that agree in their critical impact parameter and leading deflection coefficient.

\begin{acknowledgments}
We are grateful for the useful discussions with our group members. This work is supported by the National Natural Science Foundation of China (NSFC) under Grants No.12675071 and No.12175105 and NUAA funding (No. 1018-ILF26028). 
\end{acknowledgments}

\appendix

\section{Critical behavior of the Schwarzschild residual integral}
\label{app:bozza}

We examine the critical limit and turning-radius dependence of the Schwarzschild residual integral introduced in Ref.~\cite{Bozza2002}.
Following the notation of that reference, we set $x=r/(2M)$.
The metric functions are $A(x)=1-1/x$, $B(x)=A(x)^{-1}$, and $C(x)=x^2$, with $x_m=3/2$.
With $z=1-x_0/x$, the deflection integrand is $R(z,x_0)f(z,x_0)$, where $f$ is the inverse-square-root factor. 
Its approximation $f_0$ is defined by truncating $f^{-2}$ at quadratic order in $z$.
For Schwarzschild, these functions are
\begin{equation}
\begin{aligned}
 R(z,x_0)&=2,\\
 f(z,x_0)^{-2}
 &=\left(2-\frac3{x_0}\right)z
 +\left(\frac3{x_0}-1\right)z^2-\frac{z^3}{x_0},\\
 f_0(z,x_0)^{-2}
 &=\left(2-\frac3{x_0}\right)z
 +\left(\frac3{x_0}-1\right)z^2.
\end{aligned}
\label{eq:bozza_schwarzschild_roots}
\end{equation}

Before subtraction, the critical integrand and its first parameter derivative satisfy
\begin{equation}
\begin{aligned}
 (Rf)(z,x_m)&=\frac2z+O(1),\\
 \left.\frac{\partial(Rf)}{\partial x_0}\right|_{x_0=x_m}
 &=-\frac4{3z^2}+O(1),
 \qquad z\to0^+.
\end{aligned}
\label{eq:bozza_unsubtracted_derivative}
\end{equation}
Thus both the unsubtracted integrand and its first parameter derivative are nonintegrable at the lower endpoint.

After subtraction, the residual integrand $g$ and its integral $I_R$ are
\begin{equation}
\begin{aligned}
 g(z,x_0)&=2\bigl[f(z,x_0)-f_0(z,x_0)\bigr],\\
 I_R(x_0)&=\int_0^1g(z,x_0)\,\dd z.
\end{aligned}
\label{eq:bozza_residual}
\end{equation}
At the critical value,
\begin{equation}
 g(z,x_m)=\frac2z\left[
 \left(1-\frac23z\right)^{-1/2}-1
 \right]
 =\frac23+\frac z3+O(z^2),
\label{eq:bozza_critical_density}
\end{equation}
so $g(z,x_m)$ is integrable at $z=0$.
Differentiating at fixed $z$ gives
\begin{equation}
 \left.\frac{\partial g}{\partial x_0}\right|_{x_0=x_m}
 =-\frac4{3z}-\frac29+O(z).
 \label{eq:bozza_parameter_derivative}
\end{equation}
The first parameter derivative is therefore nonintegrable.
Consequently, for Schwarzschild, term-by-term integration in Eq.~(28) of Ref.~\cite{Bozza2002} does not yield a finite first-order coefficient.

The finite critical limit follows from a direct convergence argument.
For $x_0\ge x_m$ and $0<z\le1$, Eq.~\eqref{eq:bozza_schwarzschild_roots} gives $f>f_0>0$, and direct differentiation of Eq.~\eqref{eq:bozza_residual} gives $\partial_{x_0}g<0$.
Hence $0\le g(z,x_0)\le g(z,x_m)$, and dominated convergence yields, as $x_0\to x_m^+$,
\begin{equation}
 I_R(x_0)\longrightarrow b_R
 :=\int_0^1g(z,x_m)\,\dd z
 =2\ln\!\left[6(2-\sqrt3)\right].
\label{eq:bozza_finite_limit}
\end{equation}

The same monotonicity also determines whether the convergence admits a remainder of order $O(x_0-x_m)$.
For $x_0>x_m$, the difference quotient $[g(z,x_m)-g(z,x_0)]/(x_0-x_m)$ is nonnegative.
Fatou's lemma therefore gives
\begin{equation}
 \liminf_{x_0\to x_m^+}
 \frac{b_R-I_R(x_0)}{x_0-x_m}
 \ge
 \int_0^1
 \left[-\frac{\partial g}{\partial x_0}\right]_{x_0=x_m}\,\dd z
 =+\infty.
\label{eq:bozza_difference_quotient}
\end{equation}
Thus $I_R(x_0)=b_R+o(1)$, while its remainder is not $O(x_0-x_m)$.
The critical value $b_R$ agrees with Ref.~\cite{Bozza2002}, whereas the remainder estimate in its Eq.~(29) does not hold for this residual integral.

\section{Local deformations of the halo geometry}
\label{app:halo_deformation}

We construct halo geometries with the same $b_c$ and $\bar c$ but different $\ell$.
Let $A$ and $B$ denote the metric functions in Eq.~\eqref{eq:hernquist_metric}.
For a dimensionless parameter $\kappa$, we keep $C=r^2$ and define
\begin{equation}
\begin{aligned}
 \widetilde A(r)&=A(r)\ee^{\kappa w(r)},\\
 \widetilde B(r)&=B(r)+\kappa r w'(r),\\
 w(r)&=
 \begin{cases}
 (r-6)^5
 \exp\!\left[-\dfrac{(r-6)^2}{1-(r-6)^2}\right],
 &5<r<7,\\[1mm]
 0,&\text{otherwise}.
 \end{cases}
\end{aligned}
\label{eq:halo_local_deformation}
\end{equation}
We take $(a,H)=(6,9)$, for which the original metric has a quartic photon orbit at $r=6$.
The function $w$ is smooth, and all its derivatives vanish at $r=5$ and $r=7$.

Since $B=1+rA'/A$ for the original metric~\cite{Cardoso2022Halo}, the modified functions satisfy $\widetilde B=1+r\widetilde A'/\widetilde A$ and retain zero radial pressure.
The Einstein equations with $T^\mu{}_\nu=\operatorname{diag}(-\widetilde\rho,\widetilde p_r,\widetilde p_t,\widetilde p_t)$ give
\begin{equation}
\begin{aligned}
 8\pi\widetilde\rho
 =\frac{\widetilde B(\widetilde B-1)+r\widetilde B'}
 {r^2\widetilde B^2},\quad
 \widetilde p_r=0,\quad
 \widetilde p_t=\frac{\widetilde\rho}{4}(\widetilde B-1).
\end{aligned}
\label{eq:halo_deformation_source}
\end{equation}

The original density is $\rho=36(r-2)/[\pi r^2(r+6)^3]>0$, and $B>1$ for $r>2$.
By continuity on $[5,7]$, $\widetilde\rho>0$ and $\widetilde B>1$ for sufficiently small $|\kappa|$.
Thus $\widetilde p_t>0$, and the null and weak energy conditions hold throughout the exterior.
The metric remains static and regular for $r>2$.
Outside $5<r<7$, it is unchanged, preserving the horizon at $r=2$, asymptotic flatness, and the ADM mass $10M_{\rm BH}$.
The smooth matching introduces no surface layer.

The effective potential $\widetilde V=\widetilde A/r^2$ satisfies
\begin{equation}
 \frac{\widetilde V'}{\widetilde V}
 =-\frac{2(r-6)^3}{r(r-2)[(r-3)^2+63]}
 +\kappa w'(r).
 \label{eq:halo_deformation_potential}
\end{equation}
Since $w'(r)/(r-6)^3$ extends continuously to zero at $r=6$ and is bounded on $[5,7]$, sufficiently small $|\kappa|$ preserves the sign of $\widetilde V'$.
In particular, $\widetilde V$ decreases to zero for $r>6$, so returning rays can approach the quartic orbit from outside.

Near $r=6$, we have $w(r)=(r-6)^5+O((r-6)^7)$.
Consequently, $\widetilde V$ agrees with $V$ through fourth order, while $\widetilde B(6)=B(6)$ and $\widetilde B'(6)=B'(6)$.
The orbit remains at $r_c=6$, with unchanged $b_c$, $\bar c$, $S_c$, and $S_{1,c}$.
The fifth-order coefficient changes as $\widetilde h_{5,c}=h_{5,c}-\kappa A_c$, where $A_c=A(6)$.
Equation~\eqref{eq:cusp_log_coefficient} then gives
\begin{equation}
 \ell_\kappa
 =-\frac{\sqrt{2}}{5}-10368\sqrt{2}\,\kappa.
 \label{eq:halo_matched_coefficients}
\end{equation}
This construction shows that $\ell$ can distinguish geometries with the same $b_c$ and $\bar c$.

\bibliographystyle{apsrev4-2}
\bibliography{references}
\end{document}